\documentstyle [twocolumn,epsf]{mn2e}
\newcommand{\mincir}{\raise
-3.truept\hbox{\rlap{\hbox{$\sim$}}\raise4.truept\hbox{$<$}\ }}
\newcommand{\magcir}{\raise
-3.truept\hbox{\rlap{\hbox{$\sim$}}\raise4.truept\hbox{$>$}\ }}
\newcommand{\minmag}{\raise
-3.truept\hbox{\rlap{\hbox{$<$}}\raise5.truept\hbox{$<$}\ }}
\newcommand{\be}{\begin{equation}}
\newcommand{\ee}{\end{equation}}
\newcommand{\ba}{\begin{eqnarray}}
\newcommand{\ea}{\end{eqnarray}}
\newcommand{\brr}{\begin{array}}
\newcommand{\err}{\end{array}}
\newcommand{\bc}{\begin{center}}
\newcommand{\ec}{\end{center}}

\newcommand{\lb}{{\left<\right.}}
\newcommand{\rb}{{\left.\right>}}
\newcommand{\hm}{\,h^{-1}{\rm Mpc}}

\title[PSCz pdf]
{Cosmological Implications of the PSCz {\em PDF} and its Moments}
\author[Plionis \& Basilakos]
{Manolis Plionis$^{1}$ \& Spyros Basilakos$^{2}$ \\
\vspace{0.1cm}
$^1$Institute of Astronomy \& Astrophysics, National Observatory
of Athens, I.Metaxa \& B.Pavlou, Palaia Penteli, Athens 152 36, Greece \\
$^2$ Astrophysics Group, Imperial College London, Blackett Laboratory, 
Prince Consort Road, London SW7 2BW, UK\\
}

\begin{document}

\maketitle

\begin{abstract}
We compare the {\em pdf} and its low-order moments (variance and skewness) 
of the smoothed PSCz galaxy density field and of the corresponding simulated PSCz
look-alikes, generated from N-body simulations of 6 different DM models;
four structure normalized with $\Gamma=0.25$ and $\sigma_{8}=0.55 
\Omega_{\circ}^{-0.6}$, one COBE normalized and the old SCDM.
The galaxy distributions are
smoothed with a Gaussian window at three different smoothing scales
$R_{sm}=5$, $10$ and $15 \;\hm$. 
We find that the simulation PSCz look-alike {\em pdf}'s are sensitive only on the 
normalization of the power spectrum, probably
due to shape
similarity of the simulated galaxy power spectrum on the relevant scales.
We find that the only models that are consistent, at a high significance level,
with the observed PSCz {\em pdf} are models with a relatively
low power spectrum normalization $(\sigma_{8}=0.83)$. From the phenomenologically 
derived $\sigma_{8}-moments$ relation, fitted from the simulation data,
we find that the PSCz moments suggest $\sigma_{8} \simeq 0.7$.

\end{abstract}

\vspace{0.2cm}

\section{Introduction}
The study of the galaxy or cluster one-point probability density function 
({\em pdf}), 
defined as the probability of having $N$ cells of volume $V$ with overdensity 
$\delta (\equiv \rho/\langle \rho \rangle -1)$, has been used in order
to probe the Gaussian or not nature of the cosmic density field as well as cosmological
probe, since the {\em pdf} depends on the two-point correlation function or 
its Fourier Transform, the 
power spectrum, but also on higher order correlations.

The {\em pdf} of the primordial density field evolves due to gravitational instability
and even if it follows Gaussian statistics originally, it eventually
becomes non-Gaussian by the mere effects of gravity. 
One therefore may envision that smoothing on large enough scales
the present day galaxy density field one can recover the initial conditions.
This procedure is unstable due to the fact (1) that galaxies are
probably a biased tracer of the underline density field and 
(2) the central limit theorem ensures that heavy 
smoothing will Gaussianize even an inherently non-Gaussian field.
However, elaborate inversion techniques have been devised and applied with varying
level of success to real and simulation data (see Monaco et al. 2000 and references therein)
and the general outcome is that indeed the present-day galaxy distribution appears to 
originate from the effects of gravitational instability on an initial Gaussian fluctuation field.

Many authors studying the {\em pdf} statistics of different 
samples of extragalactic objects,
assuming linear biasing, have claimed
that if the large-scale distribution of galaxies trace 
the underlying matter distribution
then the observed {\em pdf} can be approximated using Gaussian initial 
conditions (cf. Bouchet et al. 1993; Gazta\~naga \& Yokohama 1993; 
Sheth et al. 1994; Kofman et al. 1994). 
Other authors, using N-body simulations, studied the 
evolution of the {\em pdf} and its moments (cf. Coles \& Frenk 1991; Juszkiewicz 
et al. 1993; Coles et al. 1993; Kofman et al. 1994; Colombi 1994) and concluded that the
normalized skewness of the matter distribution, 
$S_{3}=\langle \delta^{3} \rangle/\langle \delta^{2} \rangle$ depends 
on the shape of the power spectrum. 

Saunders et al. (1991) used the lower-order moments of the {\em pdf} (the
variance and skewness) of the sparse sampled IRAS 0.6Jy sample (QDOT)
to put constraints on cosmological models. They found that on large scales 
they are larger than the corresponding moments of QDOT
look-alikes, generated from the old standard CDM model.
In this paper we will utilize the completed PSCz-IRAS redshift survey 
in a similar attempt.

The layout of this paper is as follows: in Section 2 we describe briefly
our simulations and the considered models; in Section 3 we present the analysis
method and the results obtained; we state our main conclusions in
Section 4.

\section{The Data}
\subsection{The PSCz sample}  
In this work we use the recently completed PSCz-IRAS flux-limited 
redshift survey which is described with details in
Saunders et al. (2000). It is based on the IRAS Point Source 
Catalog and contains $\sim 15500$ galaxies with flux $S_{lim}\ge 0.6$ Jy
covering the $84\%$ of the sky. 

Due to the fact that the PSCz catalog is a flux-limited sample 
there is the well-known degradation of
sampling as a function of distance from the observer which is
represented by the selection function. It is necessary therefore
to recover the true galaxy density field from the observed 
flux-limited sample. This is done
by weighting each galaxy by $\phi^{-1}(r)$, where the selection
function, $\phi(r)$, is defined as the fraction of the galaxy number density 
that is observed above the flux limit at some distance $r$
%
In this work we use
a luminosity function of the form assumed by Saunders et al. (1990) with 
$L_{*}=10^{8.45} \; h^{2} L_{\odot}$, $\sigma=0.711$, $\alpha=1.09$ 
and $C=0.0308$ (cf. Rowan-Robinson et al. 2000). 

In order to take into account {\em (1)} the 16\% of the sky 
which is devoided of data,
due to high cirrus emission areas and unobserved regions, and 
{\em (2)} the distortion of
galaxy positions due to peculiar velocities, we utilize the 
corrected galaxy catalog of Branchini et al. (1999). These authors 
followed the Yahil et al. (1991) prescription to fill the 
galactic plane with synthetic objects that reproduce the mean density 
of galaxies of the nearby regions while at higher galactic
latitudes they fill randomly the masked regions with synthetic objects, 
reproducing the overall mean number density of PSCz galaxies.
Furthermore, in order for them to minimize the so 
called ``Kaiser'' effect (Kaiser 1987) and recover the true 
3D galaxy distances they used an iterative technique and reconstructed
the PSCz galaxy density field for different values of
$\beta=\Omega_{\circ}^{0.6}/b_{\rm IRAS}$. In our present analysis
we use their $\beta=0.5$ reconstruction, since recently Branchini 
et al. (2001),
using the VELMOD analysis to correlate the PSCz density field with
the SFI galaxy velocities (cf. Haynes et al. 1999) found 
$\beta_{IRAS} \simeq 0.42$.

\subsection{N-body Simulations}
We use mock PSCz catalogs, generated from six large cosmological N-body
simulations  (see Cole et al. 1998), in order to investigate whether the 
{\em pdf} and its moments can discriminate between models. 
The PSCz look-alikes 
have been generated by E. Branchini, choosing suitable LG-like observers
(having similar to the observed Local Group velocity, shear and
overdensity), and
he generously provided them to us. Note that due to our ignorance in
assigning galaxy formation sites to the DM halo distribution and 
for a consistent treatment of all models, a biasing factor of 1 has 
been used in generating the PSCz look-alikes.

Four of the models that we consider are structure normalized
with $\Gamma=0.25$ and fluctuation amplitude in 8 $h^{-1}$Mpc scale of
$\sigma_{8}=0.55\Omega_{\circ}^{-0.6}$ (Eke, Cole, 
\& Frenk 1996), which are: (1) a flat low-density CDM model with 
$\Omega_{\circ}=1-\Omega_{\Lambda}=0.3$ ($\Lambda_{\rm CDM1}$) 
(2) a model with $\Omega_{\circ}=1-\Omega_{\Lambda}=0.5$ ($\Lambda_{\rm CDM2}$)
(3) an $\Omega_{\circ}=1$ model ($\tau_{\rm CDM}$) and
(4) an open model with $\Omega_{\circ}=0.5$ (O$_{\rm CDM}$); a COBE normalized
Einstein de Sitter model with $\sigma_{8}=1.35$ (C$_{\rm CDM}$)
and finally the old 'standard' CDM model with $\Gamma=0.5$ and $\sigma_8=0.55$
(S$_{\rm CDM}$).

For each cosmological model we average results over
10 nearly independent mock PSCz catalogs which are treated in exactly the
same way as the real PSCz data.

\begin{table}
\caption[]{$\chi^2$ probability of significant 
differences between the model-model or
model-data pair {\em pdf}'s. Each probability value corresponds to the
pair formed between the indicated model or data in the 
first column and the first row. In bold we indicate the
results of the PSCz-model {\em pdf} comparison. For example, a high
probability value (say $\magcir 0.1$) means that the corresponding two 
{\em pdf}'s show no significant differences between them. When the
probability is less than $10^{-3}$ we write 0.00.}
\tabcolsep 3pt
\begin{tabular}{ccccccc}  \\ 
              & C$_{\rm CDM}$  & $\Lambda_{\rm CDM1}$ 
& $\tau_{\rm CDM}$ & S$_{\rm CDM}$ & O$_{\rm CDM}$ & $\Lambda_{\rm CDM2}$ 
\\ \hline
\multicolumn{7}{c}{$R_{sm}=5 \; h^{-1}$ Mpc} \\ \hline
{\bf PSCz}           & {\bf 0.00} & {\bf 0.00} & {\bf 0.00} & {\bf 0.00} & {\bf 0.09} & {\bf 0.33}\\
C$_{\rm CDM}$        &           & 1.00 & 0.00 & 0.00 & 0.00 & 0.00\\
$\Lambda_{\rm CDM1}$ &           &     & 0.00 & 0.00 & 0.00 & 0.00\\
$\tau_{\rm CDM}$     &           &     &     & 1.00 & 0.00 & 0.00\\ 
S$_{\rm CDM}$        &           &     &     &     & 0.00 & 0.00\\
O$_{\rm CDM}$        &           &     &     &     &     & 1.00 \\ \hline
\multicolumn{7}{c}{$R_{sm}=10 \; h^{-1}$ Mpc} \\ \hline
{\bf PSCz}           & {\bf 0.00} & {\bf 0.00} & {\bf 0.00} & {\bf 0.00} & {\bf 0.46} & ${\bf 0.99}$ \\
C$_{\rm CDM}$        &           & 1.00      & 0.00 & 0.00 & 0.99  & 0.04\\
$\Lambda_{\rm CDM1}$ &           &           & 0.00 & 0.00 & 0.00  & 0.00 \\
$\tau_{\rm CDM}$     &           &           &      & 0.00 & 1.00  & 0.00\\ 
S$_{\rm CDM}$        &           &           &      &      & 0.00  & 0.00\\
O$_{\rm CDM}$        &           &           &      &      &       & 1.00\\
\hline
\end{tabular}
\end{table}

\section{The {\em pdf} Derivation}
We obtain a continuous galaxy density
field by smoothing the discrete distribution of PSCz galaxies
on a $N^{3}$ grid. We use a Gaussian kernel with
smoothing radius, $R_{sm}$, varying from 5 to 
15 $h^{-1}$ Mpc. The grid size that we use is set equal to $R_{sm}$. 
If $\rho({\bf x})$ is the smoothed density field at a grid point, 
then the relative fluctuations are given by
$\delta({\bf x}) = \rho({\bf x})/\lb \rho \rb-1$, where
the average density $\lb \rho \rb$ does not depend on $R_{sm}$.
The frequency distribution of $\delta$'s is the {\em pdf}.

Basilakos, Plionis \& Rowan-Robinson (2001) investigated the 
systematic biases that enter in such a smoothing procedure due to 
the convolution between the PSCz selection 
function and the constant smoothing radius. 
\begin{figure*}
\mbox{\epsfxsize=12cm \epsffile{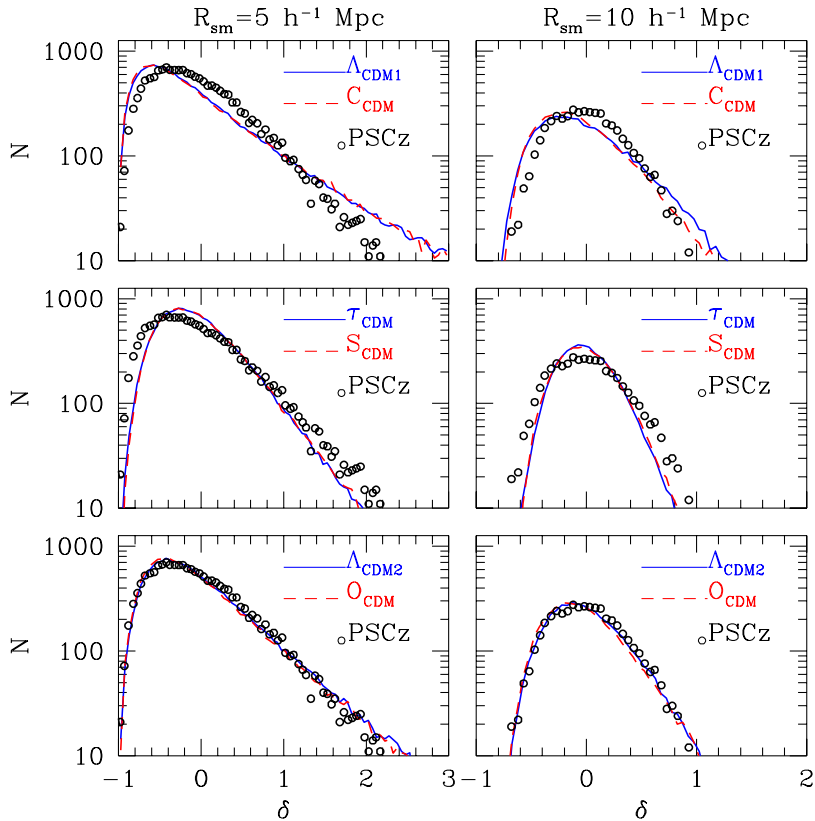}}
\caption{Comparison of the PSCz (circles) and mock {\em pdf}'s for the three different 
$\sigma_8$ families of cosmological models.}
\end{figure*}
The effect is such that the
derived smooth density field, at large distances, is overestimated in regions 
where galaxies are detected (due to the large $\phi(r)$ weighting), 
while it is underestimated in underdense regions. They
devised a phenomenological procedure that corrects effectively the high
density regions for such effects. Since this procedure is
multiplicative in nature, it is not as effective in recovering
the underdense regions and thus we confine our analysis within a limited radius,
$r_{max}$ ($=80$, 110 and 120 $h^{-1}$ Mpc for
the $R_{sm}=5$, 10 and 15 $h^{-1}$ Mpc fields, respectively), 
within which we have verified that the derived density field 
is mostly unaffected by the above mentioned bias. This was achieved
by comparing the {\em pdf} and its moments of PSCz look-alikes and the
full 3D ``galaxy'' distribution in the simulations. Even within
$r_{max}$ the raw values of the variance and skewness 
are overestimated by $\sim 20\%$, but
only for the $R_{sm}=10$ and 15 $h^{-1}$ Mpc fields.
Applying the Basilakos et al. (2001) phenomenological
correction procedure does recover the correct values of these moments.

In order to investigate how discriminative between models
is the {\em pdf}, we have performed a $\chi^2$ statistical test, 
comparing the model {\em pdf} between them. 
We find that the present {\em pdf} analysis can discriminate only 
different $\sigma_8$ models, irrespectively of the DM content
and the geometry. By using the whole 3D simulation
data, we have verified that neither shot-noise nor the PSCz 
selection function is responsible for such behaviour;
which is probably due to the shape similarity
of the derived ``galaxy'' power spectrum on the relevant scales 
(see figure 2 of Cole et al. 1998). This is particularly true for the 
O$_{\rm CDM}$ and $\Lambda_{\rm CDM2}$ models that both have 
$\Gamma=0.25$ and $\sigma_8=0.83$. 

Furthermore, we find that the most effective, in discriminating power,
smoothing scale is the $R_{sm}=5$ $h^{-1}$ Mpc one.
In Table 1 we present the probabilities, $P_{>\chi^{2}}$, 
of significant differences between any two {\em pdf}'s for the two indicated
smoothing scales. It is clear that the statistical test
identifies 3 families of consistent among them models, all of which are 
based on their relevant $\sigma_{8}$ normalization. 

In figure 1 we present a comparison between the PSCz {\em pdf} and 
those of the 3 families of models. From this figure as well as from table 1, 
it is clear that the models with $\sigma_{8}=0.83$ are the only ones 
consistent with the data.

\subsection{Moments of the {\em pdf}}
The variance $\sigma^2$ and the skewness $\gamma$ are
defined as the second-- and third-- order moments of the 
$\delta$ field, respectively. We therefore have
\be
\sigma^2=\frac{1}{n}\sum^{n} \delta^2({\bf x})
\ee
\be
\gamma=\frac{1}{n}\sum^{n} \delta^3({\bf x}) \,,
\ee
where $n$ is the total number of grid points within $r_{max}$. Gazta\~naga \& 
Yokoyama (1993) have shown that the smoothing process itself suppresses
considerably the shot--noise effects. 
A proper treatment of these effects should entail
subtracting the ``Poisson'' terms from $\sigma$ and $\gamma$
but in conjunction with the effects of smoothing, which is not a
trivial task. However, since all the mock PSCz distributions have the
same $\langle \rho \rangle$ and we treat them similarly, the possible 
shot-noise effects are {\em relatively} and {\em qualitatively} cancelled out in
the model and data intercomparison.
\begin{table}
\caption[]{Moments of the smoothed density distribution of the mock
and real PSCz data. The uncertainties of the models are estimated from 10 
PSCz look-alike realizations of each model while those of the PSCz data 
are estimated by dividing the available volume into 4 independent regions.}
\tabcolsep 10pt
\begin{tabular}{cccc}  \\ \hline
Model & ${\bf R_{sm}}$ & ${\bf \sigma^{2}}$ & ${\bf \gamma}$  \\ \hline
{\bf C}$_{\rm CDM}$       & 5 &0.839$\pm$0.135 &2.342$\pm$1.256 \\
$\sigma_8=1.35$  & 10 &0.170$\pm$0.014 &0.073$\pm$0.031 \\
                 & 15 &0.061$\pm$0.007 &0.010$\pm$0.006 \\
${\bf \Lambda}_{\rm CDM1}$      & 5 &0.856$\pm$0.098 &2.589$\pm$1.111 \\
$\sigma_8=1.13$         & 10 &0.210$\pm$0.028&0.127$\pm$0.054 \\
                    & 15 &0.083$\pm$0.010 &0.020$\pm$0.009 \\
{\bf O}$_{\rm CDM}$ & 5 &0.477$\pm$0.067 &0.729$\pm$0.355 \\ 
$\sigma_8=0.83$ & 10 &0.129$\pm$0.014 &0.047$\pm$0.027 \\
                & 15 &0.054$\pm$0.006 &0.009$\pm$0.008 \\
${\bf \Lambda}_{\rm CDM2}$      & 5 &0.475$\pm$0.065 &0.723$\pm$0.295 \\
$\sigma_8=0.83$         & 10 &0.125$\pm$0.017&0.045$\pm$0.023 \\
                   & 15 &0.062$\pm$0.037 &0.017$\pm$0.031 \\
{\bf S}$_{\rm CDM}$ & 5   &0.263$\pm$0.025 &0.187$\pm$0.063 \\
$\sigma_8=0.55$  & 10 &0.074$\pm$0.005 &0.015$\pm$0.005 \\
                 & 15 &0.033$\pm$0.003 &0.005$\pm$0.003 \\
${\bf \tau}_{\rm CDM}$ & 5 &0.261$\pm$0.021 &0.170$\pm$0.042 \\ 
$\sigma_8=0.55$ & 10 &0.068$\pm$0.006 &0.014$\pm$0.005 \\
                & 15 &0.029$\pm$0.004 &0.003$\pm$0.003 \\
{\bf PSCz} & 5 &0.361$\pm 0.033$ &0.326$\pm 0.188$ \\
           & 10 &0.109$\pm 0.023$ & 0.025$\pm 0.012$ \\
           & 15 & 0.048$\pm 0.014$  & 0.003$\pm 0.002$\\ \hline
\end{tabular}
\end{table}

In Figure 2 we plot the $\sigma^2$--$\gamma$
plane, following Coles \& Frenk (1991), for the 
PSCz density field (red filled dots) and for the simulation mock PSCz fields.
The results of the $R_{sm}=5$, 10 and 15 $h^{-1}$ Mpc field are represented 
as open circles, triangles and squares, respectively.
Each simulation point represents one of the 10
realizations, so that the scatter represents the effect of
cosmic variance. The PSCz errorbars represent the scatter 
estimated by dividing the whole available volume in 4 equal and
independent subsamples.

We can estimate the reduced skewness by least-square fitting 
a relation: $\log \gamma =\log(\sigma^2)^\alpha + \log S_3$, 
which for $\alpha=2$ is a generic 
prediction of perturbation theory, of the hierarchical 
and log-normal clustering models on linear scales (cf. Coles \& Frenk 1991). 
According to
Bouchet et al. (1992) the reduced moments are independent of the density
and geometry of cosmological models and for this reason they have been used 
to discriminate between different biasing models rather than
cosmological models (cf. Gazta\~naga \& Frieman 1994). 
Indeed, keeping $\alpha=2$ we find $S_{3} \simeq 2.7$ for all the models, 
with scatter over the 10 realizations, of $\delta S_3 \simeq 0.4$. This
value is
in agreement, within the errors, with the PSCz value ($S_{3}^{PSCz}
\simeq 2.1 \pm 0.6$), where the scatter is from the 4 independent
volumes used in the estimate. Furthermore, this value is 
in excellent agreement with $S_3 \simeq 2.2$, estimated from the
IRAS 1.2Jy sample (cf. Fry \& Gazta\~naga 1994).
Leaving unconstrained the value of $\alpha$ we find, for both PSCz data
and models, which by construction have Gaussian initial conditions,
$S_3$ values in the range $3.1 - 3.6$ and $\alpha\simeq
2.2\pm 0.15$. The deviation from $\alpha=2$ is probably due to the
effects of shot-noise.

These results argue against large-scale
non-gaussianity in the Universe since, according the study of
Coles et al (1993), non-gaussian models show departures from such a 
$\sigma^2-\gamma$ relation.

In Table 2 we report the real and simulation PSCz {\em pdf} moments. 
Again from both the figure 2 and table 2 it is evident that the PSCz results 
are nearer to the $\sigma_8=0.83$ simulations.
\begin{figure}
\mbox{\epsfxsize=8.8cm \epsfysize=9.5cm \epsffile{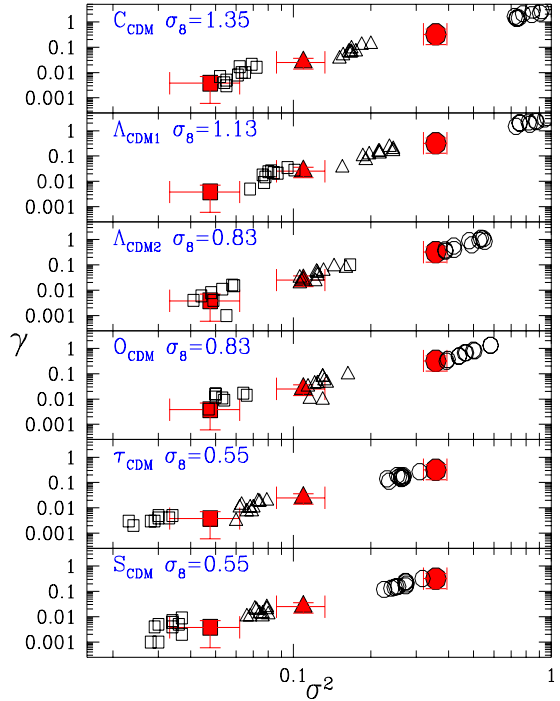}}
\caption{Scatter--plots for the variance--skewness relation. Each point
refers to a single realization of the corresponding
model. Squares, triangles and circles are for $R_{sm}=5$, 10 and 15$\hm$
respectively. Large filled symbols represent the real PSCz moments.}
\end{figure}

\subsection{Recovering the $\sigma_{8}$ of our Universe}
What is also evident, from figure 2, is that there is a tight scaling between
$\sigma_{8}$ and the values of $\sigma^2$ and $\gamma$ in the simulation 
PSCz look-alike fields. We have verified this by fitting
a linear relation to the robust $R_{sm}=5$ $h^{-1}$ Mpc result, to find
$$\sigma_8 = 1.062 (\pm 0.06) \sigma^2 + 0.313 (\pm 0.03)$$
$$\sigma_8 = 0.189 (\pm 0.02) \gamma + 0.661 (\pm 0.03)$$
with a correlation coefficient of $R=0.93$. 

Since from the simulation data we have seen that the {\em pdf} 
moments depend only on the value of $\sigma_8$, we can now 
attempt to recover, by inserting our PSCz moments to the 
above relations, the $\sigma_8$ of our Universe. We find 
$$\sigma_{8} \simeq 0.7 \pm 0.03 \;\;,$$ 
where the quoted uncertainty is
the scatter of the mean from using the raw or corrected moments 
(see section 3) and the $\sigma_8$-variance or skewness relation.
This result is in agreement with many 
recent studies (cf. Eke et al 1998; Einasto et al. 1999;
Moscardini, Matarrese \& Mo 2001) and in particular with the most
recent cluster X-ray luminosity function evolution results from the 
ROSAT Deep Cluster Survey (Borgani et al 2001).

\section{Conclusions}
We have derived the {\em pdf} moments of the PSCz galaxy catalog and
compared them with the corresponding moments of mock PSCz distributions, 
generated from six different cosmological models.
The {\em pdf} and its low-order
moments were estimated from the Gaussianly smoothed density field on a
grid, using three different smoothing scales, $R_{sm}=5$, 10 and 15$\hm$. 
For each model we have used 10 PSCz-like realizations 
to have a handle of cosmic variance. We find that this type of analysis
can discriminate models only on the basis of their 
different $\sigma_{8}$ values. We find that from the three $\sigma_8$
families of CDM models ($\sigma_8=0.55$, 0.83 and $>1.1$) the 
PSCz {\em pdf} and its
lower order moments are consistent with models having intermediate
values of $\sigma_{8}(=0.83)$.

Furthermore, using a phenomenological $\sigma_{8}-moments$
relation, derived from the model PSCz look-alike {\em pdf}'s, we find that the
PSCz data strongly suggest $\sigma_{8} \simeq 0.7 \pm 0.03$.

\section*{Acknowledgments}
The authors wish to thank E.Branchini for providing us with his reconstructed 
PSCz density field, with the PSCz simulation look-alikes and for useful
discussion.  M.Plionis 
acknowledges the hospitality of the Astrophysics Group of Imperial College, 
where this work was completed.

\end{document}